# Free-ranging dogs quickly learn to recognize a rewarding person

Srijaya Nandi[a], Mousumi Chakraborty[a], Aesha Lahiri[a], Hindolii Gope[a], Sujata Khan Bhaduri[b,a], Anindita Bhadra[a*]

**Affiliations**

a) Indian Institute of Science Education & Research (IISER) Kolkata, Mohanpur, Nadia, West Bengal, India, PIN: 741246

b) Indian Institute of Science Education & Research (IISER) Mohali, Knowledge City, Sector-81, Mohali, Punjab, India, PIN: 140306

\* Corresponding Author.

Email address: abhadra@iiserkol.ac.in (A. Bhadra).

**Abstract**

Individual human recognition is important for species that live in close proximity to humans. Numerous studies on domesticated species and urban-adapted birds have highlighted this ability. One such species which is heavily reliant on humans is the free-ranging dog. Very little knowledge exists on the amount of time taken by free-ranging dogs to learn and remember individual humans. Due to their territorial nature, they have a high probability of encountering the same people multiple times on the streets. Being able to distinguish individual humans might be helpful in making decisions regarding people from whom to beg for food or social reward. We investigated if free-ranging dogs are capable of identifying the person rewarding them and the amount of time required for them to learn it. We conducted field trials on randomly selected adult free-ranging dogs in West Bengal, India. On Day 1, a choice test was conducted. The experimenter chosen did not provide reward while the other experimenter provided a piece of boiled chicken followed by petting. The person giving reward on Day 1 served as the correct choice on four subsequent days of training. Day 6 was the test day when none of the experimenters had a reward. We analyzed the choice made by the dogs, the time taken to approach during the choice tests, and the socialization index, which was calculated based on the intensity of affiliative behaviour shown towards the experimenters. The dogs made correct choices at a significantly higher rate on the fifth and sixth days, as compared to Day 2, suggesting learning. This is the first study aiming to understand the time taken for individual human recognition in free-ranging dogs, and can serve as the scaffold for future studies to understand the dog-human relationship in open environments, like urban ecosystems.

**Keywords**

Free-ranging dogs, learning, human recognition, human-dog interaction

1. Introduction

Learning is a phenomenon that involves a long-term change in memory and/or behaviour in animals. It involves the internal representation of newly acquired information from internal and external environments and the development of new motor patterns (Dukas, 2013). The first formal experiments on learning were carried out by the Russian physiologist Ivan Pavlov on dogs (Pavlov, 1910). Ever since his discovery of classical conditioning, there have been numerous studies trying to understand learning across taxa. Learning is prevalent in almost all animal taxa, from invertebrates (Fiorito and Scotto, 1992; Hawkins and Byrne, 2015) to vertebrates. It enables the animal to make decisions that have fitness consequences, helping organisms to adjust to novel and rapidly changing environments (Baldwin 1896; Morgan 1896; Osborn 1896; Robinson & Dukas 1999).

One of the most widely encountered and advantageous forms of learning in nature is social learning, which involves the recognition of conspecifics and heterospecifics. Recognition of other individuals of their kind, also termed conspecific recognition, is important in social species, for efficient transfer of information and maintenance of social integrity of the group (Tibbetts and Dale, 2007). Intraspecific recognition is also important for guiding preferential care towards mates and kin (Marzluff and Balda 1992), reducing the intensity of fights with known competitors (Brooks & Falls 1975) and maintaining stable social bonds (Barnard & Burk 1979), all of which have fitness benefits. Heterospecific recognition or recognition of members of different species is also equally important in the context of prey-predator recognition (Mc Lean et al., 2000) and mixed species groups (Aplin et al., 2012; Mukherjee and Bhat, 2023), where identification of a species in general is more advantageous than individual recognition. Individual identity and recognition within a social group is widely known, and has multiple functions. It is a faculty that helps to define social roles, positions in social hierarchies, mate choice, recognition of kin, etc. Individual recognition across species is relatively rare, but if present, can be useful for animals that have close associations with individuals of another species. One of the examples of individual recognition is between animals and humans. This is particularly important for domesticated species especially those kept as pets and farm animals and being taken care of by human handlers.

When it comes to animals kept as pets, individual recognition of their owners is of utmost importance as the animals are entirely dependent on their owners for food and care. Domestic cats have been demonstrated to be able to distinguish between their owners and strangers by vocal cues alone (Saito and Shinozuka, 2013). Adult domestic dogs are capable of distinguishing between their owners and strangers and have been shown to exhibit patterns of attachment behaviour towards their owners (Topa´l et al., 1998). Dogs are also capable of matching their owner's voices and faces (Adachi et al., 2007).

Social learning in farm animals has been quite extensively studied, which looks into the time required for human recognition and social bond formation. The study conducted by Miura et al. (1995) on weaning pigs concluded that interaction of more than 2 weeks is essential for establishing a firm bond between pigs and their human handlers. Munksgaard et al. (2001) in their study on dairy cows found three interactions to be sufficient in eliciting recognition between a gentle and aversive handler. This recognition was displayed by the tendency of the cows to avoid the aversive handler and approach closer to the gentle handler. Latency to approach, which can be defined as the time taken by an animal to approach close to the subject under consideration, has often been used as an indicator of familiarity. Koba and Tanida (1999) working with a group of pigs found that the latency of contact with a handler declined significantly after just 2 days of positive interactions.

Individual human recognition is also found in urban-adapted species that inhabit a human-dominated landscape and exploit humans as resources. Stephan et al. (2012) found pigeons to be successful at using facial cues to distinguish between familiar and unfamiliar humans. Northern mockingbirds were observed to be capable of quickly identifying threatening intruders near their nest after just two interactions consisting of 30s each (Levey et al., 2009). American crows quickly learn and recognize a dangerous person and successfully retain the memory for at least 2.7 years (Marzluff et al., 2010). Wild magpies are also capable of distinguishing between threatening and non-threatening humans wearing same-coloured dresses (Lee et al., 2011).

Domestic dogs (*Canis lupus familiaris)* have an interesting history of evolution through domestication from wild wolf-like ancestors. Dogs are a species that share a unique bond with humans, and have thus been studied extensively for various social cognitive abilities, that enable them to bond with humans better. Dogs that live among humans as pets are capable of recognizing their humans, not only in person, but also from their voices and photographs (Adachi et al., 2007; Eatherington et al., 2020). However, among the canids, the domestic dogs are not unique in their ability to recognise individual humans. Wild gray wolves (*Canis lupus lupus*) in captivity are capable of discriminating between familiar and unfamiliar human voices (Gammino et al., 2023). This indicates that domestication is not a prerequisite for the ability of individual human recognition but might have enhanced the ability in the dogs. Free-ranging dogs are ubiquitously found in human habitats across the Global South, and though they don't have owners, they constantly interact with humans in multiple ways. While some humans are a source of food, shelter and even care for them, others can be threats, as people's reactions towards free-ranging dogs range from extremely positive to extremely negative. It is thus interesting to understand the extent to which free-ranging dogs are capable of recognizing individual humans, as this might be impacting their survival in the human-dominated landscape.

In India, free-ranging dogs have lived as continuous populations for centuries (Thapar, 1990) and are present in almost all human habitations in rural to urban areas (Vanak and Gompper, 2009a). They are primarily scavengers, dependent on human-generated waste for their survival, but also actively beg from people and sometimes hunt. They often find shelters in and around public as well as private spaces like parks, garages, backyards, etc. They scatter garbage, defecate in the open, disturb people with their nocturnal barking and act as a reservoir of zoonotic diseases like rabies (Fekadu, 1982), thereby often acting as a nuisance. There are cases of dogs chasing or attacking people and also of killing wildlife, because of which a part of the human population is extremely antagonistic to the free-ranging dogs (Agarwal and Reddajah, 2004; Butler et al, 2004; Fekadu, 1982; Corfmat et al., 2023). On the other hand, they are regularly fed and cared for by some. However, incidents of dogs getting beaten and killed by humans are also not scarce and humans are a major factor for early-life mortality of these dogs (Paul et al., 2016). The interactions they receive from humans largely influence their behaviour towards humans, and the FRDs exhibit substantial behavioural plasticity. Pups have been found to readily follow human pointing gestures while juveniles do not. Adult dogs rely on reliability cues to adjust their responses to human pointing (Bhattacharjee et al., 2017b). Therefore, it is adaptive for dogs to understand the intentions of unfamiliar humans before interacting with them. Canines have been suggested to have the propensity to attend to the actions of their social companions and learn to recognize humans as companions and understand the relationship through learning and experience (Reid, 2009). Also, social attachment with humans is a key driver of the process of domestication (Nagasawa et al., 2015). Since they receive a myriad of interactions from humans on a daily basis, it would be advantageous for them to have the capability of individual human recognition. Retaining the memory of prior interactions with individual humans will enable them to take decisions about how to interact with them in subsequent encounters.

Free-ranging dogs have been reported to show reduced latency to approach an initially unfamiliar person providing petting during repeated interactions (Bhattacharjee et al., 2017). However, this study did not look for the capability of free-ranging dogs to distinguish the experimenter from another individual who did not give any reward. Thus, we cannot be sure if the dogs actually had the capacity of individual recognition, or were showing reduced latency due to the familiarity to the experimental paradigm. We carried out an experiment to address this question, presenting free-ranging dogs with two individuals, one who gave a reward and the other who did not. Over repeated trials, we tested the ability of the dogs to choose the target individual with lowered latency and higher socialization index, which was considered as a marker for learning.

## 2. Materials and methods
### 2.1. Subjects

A total of 53 randomly selected free-ranging dogs (27 males and 26 females) were used in the study. All the dogs were adults i.e., greater than or equal to one year of age. To ensure that the dogs were adults, only males whose testes had descended and females whose nipples appeared dark greyish to black were selected. The age of the dogs studied were also confirmed by speaking to local residents. Only visibly healthy dogs showing no signs of injury like wounds on their bodies or having difficulty in walking were selected. This was done to reduce the chance of the dogs dying within seven days of the experiment and to ensure that poor health conditions did not affect their response during the experiments.

### 2.2. Study areas

The experiments were conducted in Kalyani (22°58′30″N, 88°26′04″E), Kalyani Simanta (22°59'14.41"N, 88°25'44.39"E) and Anandanagar (22°58'32.10"N, 88°29'5.26"E) in Nadia district of West Bengal, India (Supplementary Fig 1).

### 2.3. Experimenters involved

Four female experimenters almost of the same age and unfamiliar to the dogs studied participated in the experiments. Experimenter 1 (E1) and experimenter 2 (E2) were the primary experimenters between whom the dogs chose during the concurrent choice tests. They were of a similar height and build. Experimenter 3 (E3) acted as the camera person who recorded the experiments and kept track of time. Experimenter 4 (E4) participated only during the approachability test conducted on Day 0. The entire experiment was video recorded by E3 on all days.

### 2.4. Experimental procedure

**Approachability test**

The approachability test was conducted to understand the sociability in dogs which can be defined as their tendency to approach an unfamiliar person (Bhattacharjee et al., 2021). On the first day of the experiment (Day 0), the experimenters walked in an area, and on sighting one or more dogs, they performed this test on individuals, rather than a group of dogs. E4 stood at an approximate distance of 4-5 m from the focal dog and called out using a positive vocalization "ae-ae-ae" for a maximum duration of 60s. This vocalization is very commonly used by people of West Bengal to seek attention of dogs (Bhattacharjee et al., 2017). Since the dogs were free-ranging and not on leash, E4 had to move in order to adjust her position with respect to the dog using eye estimation. E4 also gazed at the dog throughout the experiment. The dogs were given a maximum of 60s to approach within 1 body length (approximately 0.8m) of E4 (Fig 1., ESM Video 1). Only dogs that approached within 0.8m (one body length of the dog) from E4 within 60s and did not show any signs of fear (like tail droop, crouched body, etc.) or aggression (like barking, growling, snarling, etc.) on Day 0 were selected for further experimentation. The selected dogs were photographed and their locations were marked using the GPS application of cellular phones for tracking purposes.

**Learning and test phase**

The learning phase involved the period during which the dogs became aware of the identity of the person from whom they received rewards. The test phase involved a concurrent choice test where the choice of the dogs was recorded. An experimenter was considered to be chosen if the dog being tested approached within one body length of the experimenter and at least gazed at her (ESM video 2). The other behaviours shown by the dogs and the time taken by the dogs to approach within one body length of the experimenter were also noted. Both the learning and test phases were conducted from Day 1 to Day 5 while only the test phase was conducted on Day 6.

**Day 1:** The experiment began with conducting a concurrent choice test (test phase) on Day 1. E1 and E2 stood in front of the dog at a distance of approximately 1-1.5 m from each other and 4-5 m from the dog. The focal dog, E1 and E2 stood in an arrangement that can be imagined as the vertices of an isosceles triangle (Fig 1). Both the experimenters held a piece of boiled chicken in their right hand and folded both of their hands behind their back to ensure that the chicken piece was not visible to the dog. They gazed (made eye contact) at the dog and called it using the sound "ae-ae-ae" for a maximum duration of 60 seconds. The choice made by the dog on Day 1 determined which experimenter (E1 or E2) would provide the reward to that particular dog on the four subsequent days of the experiment. The individual that the dog chose on Day 1 did not provide any reward. After the dog made its choice, the experimenter chosen stopped calling and making eye contact with the dog (looked away) and did not give it any reward. The other experimenter whom the dog did not choose kept calling the dog while gazing at it. After the dog approached within one body length of the other experimenter, the dog was given the reward (learning phase). The individual who gave the reward on Day 1 gave the reward to that particular dog on all subsequent days and also acted as the correct choice/person for that particular dog. The other experimenter whom the dog chose first on Day 1 and who did not give reward to the dog acted as the incorrect choice/person for the subsequent days (Figure 1). The reward given was a piece of boiled chicken weighing approximately 8-10g followed by 10s of petting. The dogs were petted on their head and neck by the experimenters using their right hand (ESM video 2).

**Days 2- 5:** From Day 2 to Day 5, if the dog chose the correct person, the dog was immediately given the reward. However, if it chose the incorrect person, the dog was denied a reward from her and was subsequently called by the correct person and given the reward.

**Test phase:**

On Day 6, a trial was conducted where none of the experimenters had food with them (but held their hands behind their back just like the previous days) and the dog had to choose between them (ESM video 3).

To reduce the preference towards the experimenters which might arise due to differences in the colour of the dresses worn, both E1 and E2 wore identical dresses (same coloured trousers, shirts and shoes). The side on which the two experimenters (E1 and E2) stood on the six days of the experiment (Day 1 to Day 6) was randomized to avoid the chances of the dogs learning the side from which they received the reward rather than the person. All the experiments were conducted between 14:00 h and 18:00 h.

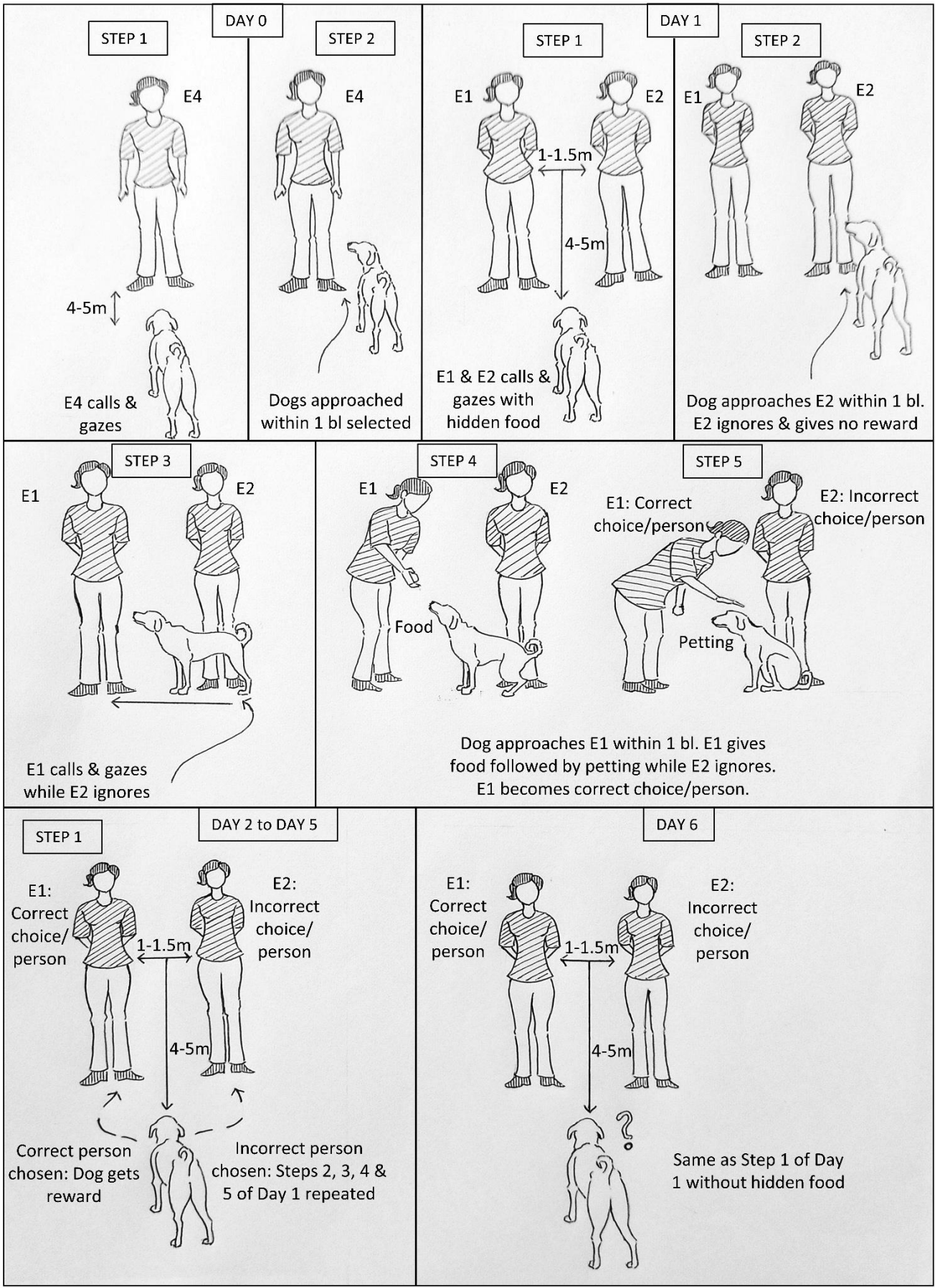

**Fig 1.** Schematic representation of the experimental procedure. Abbreviations: a) E1 and E2: Experimenter 1 and Experimenter 2, b) bl: body length of dog, which is the length from the nose tip to the tail base (1bl = ~0.8m). (Credit: Arpan Bhattacharyya)

### 2.5. Data collection and analysis

All the videos were coded by S.N., and the obtained data was used for further analysis. Only the dogs that completed all six trials (up to Day 6) were used for analysis. We checked for three parameters across different days of the experiment which were namely choice, latency of approach (in seconds) and socialization index (SI).

A choice was scored if a focal dog approached close to i.e. within one dog body length of the experimenter and at least gazed at her during the concurrent choice tests. The latency of approach was measured in seconds and was defined as the time taken by the dogs to approach within one dog body length after the dog noticed the experimenters for the first time. The socialization index (SI) was constructed following the method used by Bhattacharjee et al. (2017), with a few modifications (Table 1). Different behaviours shown by the dogs were scored based on the energy expended for performing the behaviours, the level of risk associated and the level of affiliation involved (based on the personal observation of the experimenters). The behaviours displayed by the dogs towards the experimenters while they were close to 1 body length from the experimenter and before receiving any cue (reward from the correct person or gazing away from the incorrect person) from them were used for calculating SI. Tail wagging by free-ranging dogs is a sign of affirmative behaviour and it has been shown that they show frequent tail wagging along with gazing when faced with an unfamiliar solvable task (Bhattacharjee et al., 2017a). Tail wagging is also used by dogs to show affiliation towards humans and is indicative of a positive social bond (Bhattacharjee et al., 2017). No tail wag was given a score of 0 since it required no energy expenditure and also showed no dog-initiated affiliation. Jumping and affiliative vocalization, on the other hand, were given a score of 6.

Jumping was given a high score since it required a substantial amount of energy and had a high associated risk of receiving a negative interaction from humans (personal observations found that most people do not like free-ranging dogs to physically contact them). Affiliative vocalization was also given a high score since it is performed very rarely, only when the dogs are trying to be very affiliative with humans. Licking, pawing and nudging (involving physical contact with the experimenters) were given intermediate scores since they involved a moderate amount of associated risk and energy expenditure.

| Behaviour | Score |
|---|---|
| No tail wag | 0 |
| Slow tail wag | 1 |
| Fast tail wag (without back movement) | 2 |
| Rapid tail wag (with back movement) | 3 |
| Licking experimenter | 4 |
| Pawing experimenter (using either the left or right paw to touch the experimenter | 5 |
| Jumping on experimenter's body/Affiliative vocalizations while looking at experimenter | 6 |

**Table 1.** Behaviours shown by free-ranging dogs and their associated scores used to calculate the SI.

Latency of approach and socialization index (SI) were checked for normality using the Shapiro-Wilk test. Both these parameters were not normally distributed and therefore non-parametric tests were performed.

Latency of approach and SI for males and females during the approachability test was compared using the Mann-Whitney U test. A Chi-square goodness-of-fit test was performed to check the preference of dogs for the two experimenters (E1 and E2) participating in the choice tests. To understand the days on which the choice made by the dogs differed from chance, a Chi-square goodness of fit test was performed. A Chi-square test of independence was performed to check if the day of the experiment influenced the number of correct and incorrect choices made by the dogs. A post hoc test with Bonferroni correction was performed to compare between the number of correct and incorrect choices on the different days. To compare latency of approach and SI of first choice across days, a Friedman test was performed. A post hoc Wilcoxon rank sum test with Bonferroni correction was conducted for pairwise comparison between the days. To compare the SI for the correct versus incorrect person (only for dogs that made an incorrect choice, i.e. chose the incorrect person first followed by approaching the correct person), the Wilcoxon signed rank test was performed for each day. A Mann-Whitney U test was performed to check if the latency of approach and SI differed between the dogs that chose an incorrect person as opposed to dogs that chose the correct person which was also performed for each day separately.

A generalized linear mixed model (GLMM) with binomial distribution was performed to check the effect of day of the experiment, latency of approach and SI on the choice (correct or incorrect) made. The identity of individual dogs was included as a random effect in the intercept. AIC values were used for selecting the best fitting model. The GLMM was performed using the lme4 package. All statistical analyses were conducted using R Studio (R Development Core Team, 2022). An alpha level of 0.05 was used for all the analysis conducted.

### 3. Results

53 of the 110 dogs tested responded in the approachability test, and were used for further trials in the learning and test phases. 45 dogs successfully completed all six trials and were considered for the analysis reported henceforth.

#### 3.1. Latency of approach during the approachability test

The latency of approach during the approachability test conducted on Day 0 was (12.44 ± 11.88) s with a median of 7s. Male and female dogs did not differ significantly in their latency of approach towards an unfamiliar person (Mann-Whitney U test: W=271.5, *P*=0.682, Fig 2a).

### 3.2. SI during the approachability test

The SI during the approachability test conducted on Day 0 was 2.09 ± 1.63 with a median of 2. Male and female dogs did not differ significantly in their SI towards an unfamiliar person (Mann-Whitney U test: W=272, *P*=0.636, Fig 2b).

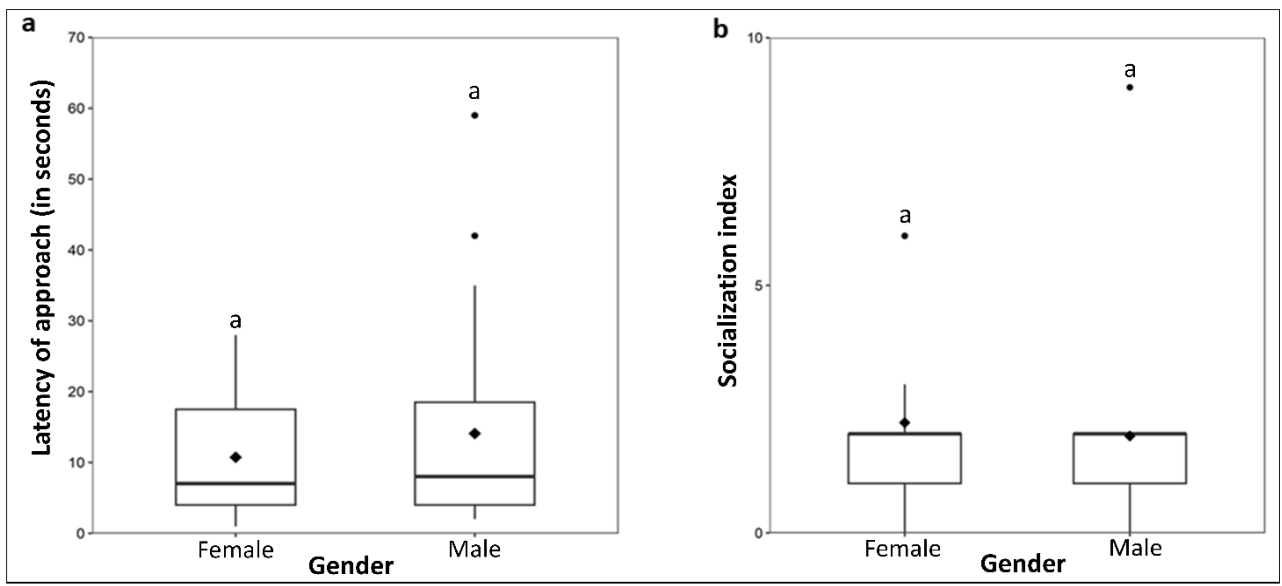

**Fig 2.** A box and whiskers plot showing (a) Latency of approach (in seconds) and (b) Socialization index (SI) for male and female free-ranging dogs towards an unfamiliar person during the approachability test conducted on Day 0. Rhombus indicates the mean. The thick horizontal black line passing through the boxplots indicates median and the dots indicate the outliers.

### 3.3. Choice between experimenters on Day 1

Out of the 45 trials conducted on Day 1, E1 was chosen 26 (57.78%) times while E2 was chosen 19 (42.22%) times. There was no significant difference in the preference of dogs for the two experimenters (Chi-square goodness-of-fit test: $\chi^2$=1.0889, df=1, *P*=0.297, Supplementary Fig 2).

### 3.4. Correct choice across days

The percentage of correct choice was significantly higher than chance only on Day 5 (Chi-square goodness-of-fit test: χ2=11.756, df=1, *P*=0.0006) and Day 6 (Chi-square goodness-of-fit test: χ2=8.022, df=1, *P*=0.005). For comparison between correct and incorrect choice on all days (from Day 2 to 6), check Supplementary Table 1.

An association was found between the day of the experiment and the number of correct and incorrect choices made (Chi-square test of independence: χ2= 12.856, df=4, *P*= 0.012). A post-hoc test with Bonferroni correction (adjusted α=0.0125) indicated Day 5 (χ2= 8.99, df=1, *P*= 0.003) and Day 6 (χ2= 6.516, df=1, *P*= 0.011) to have significantly higher correct choices when compared to Day 2 (Fig 3). For comparison between correct and incorrect choices between Day 2 and all the remaining days (Day 3 to 6), check Supplementary Table 2.

Based on the AIC values, the best-fitting model depicted both SI of first choice or SI-1 (where first choice is the person chosen first) and the day of the experiment to be significant predictors of the dog's ability to make a correct or incorrect choice. SI-1 is explained below.

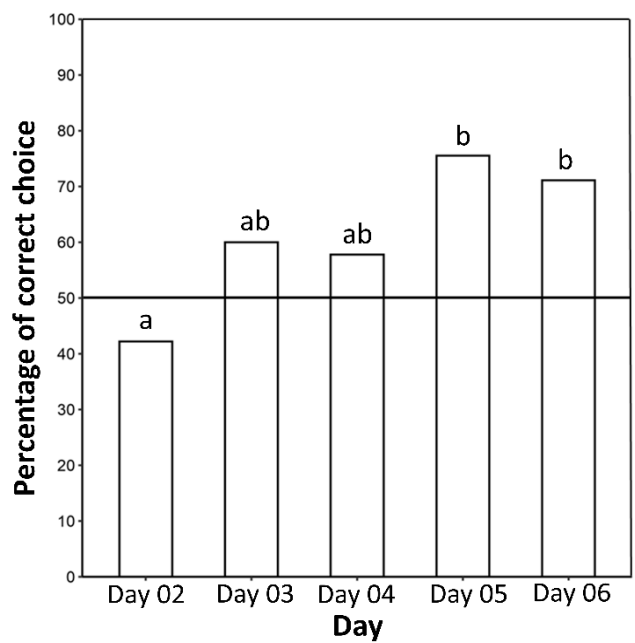

**Fig 3.** A bar graph showing the percentage of correct choice (correct choice is equivalent to the correct person chosen) from Day 2 to Day 6.

### 3.5. Latency of approach for first choice

The latency of approach for first choice (latency-1) was defined as the time taken by the dogs to approach the first experimenter who might be the correct or the incorrect person. Latency-1 was found to differ significantly across days (Friedman chi-squared = 46.001, df = 5, $P$ = 9.079e-09). Latency-1 declined significantly from Day 1 to Day 2 (post-hoc pairwise Wilcoxon rank sum test with Bonferroni correction: $P$ = 0.022, Fig 4a), following which it showed saturation. The latency-1 was also significantly lower on all days when compared to Day 1. For details on the $P$ values, check Supplementary Table 3.

### 3.6. SI during first choice

The socialization index during first choice (SI-1) was calculated based on the kind of behaviours the dogs showed towards the first person whom they chose in the concurrent choice test, irrespective of the choice being correct or not. SI-1 was found to differ significantly across days (Friedman chi-squared = 64.699, df = 5, p-value = 1.294e-12). The SI-1 significantly increased from Day 1 to Day 3 (post-hoc pairwise Wilcoxon rank sum test with Bonferroni correction: $P$=0.0004, Fig 4b). The SI-1 was also significantly higher than Day 1 on all days except Day 2. For details on the $P$ values, check Supplementary Table 4.

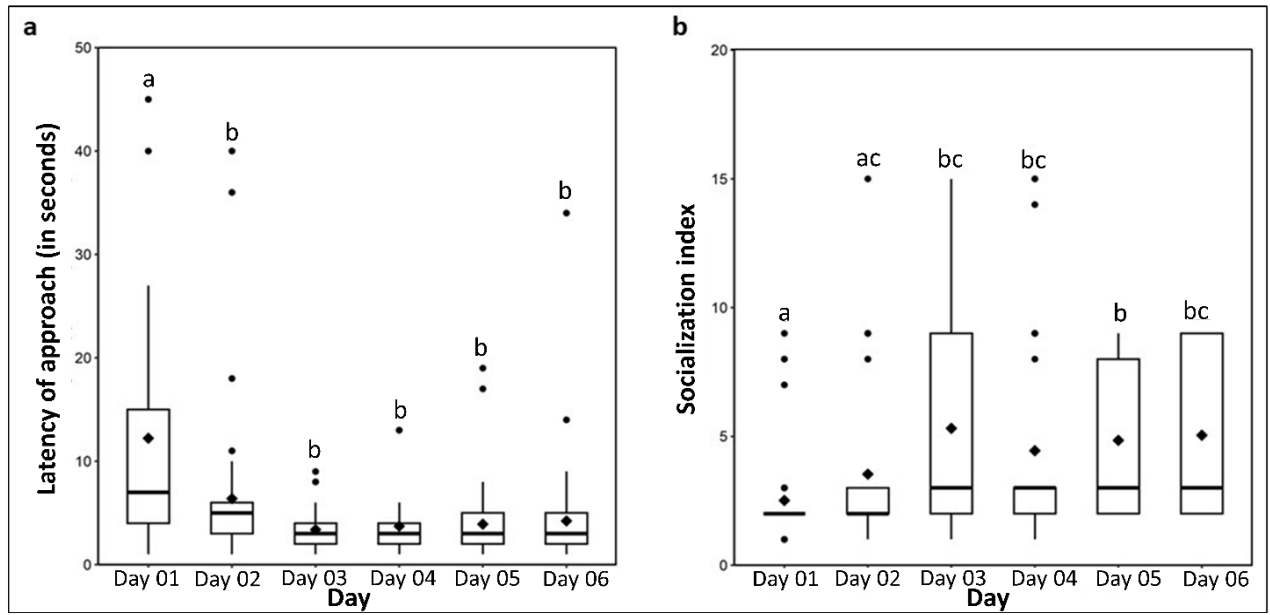

**Fig 4.** A box and whiskers plot showing (a) Latency of approach (in seconds) during first choice (latency-1) which is defined as the time taken to approach the person chosen first, irrespective of the choice being correct or incorrect and (b) Socialization index (SI) during first choice (SI-1) which is defined as the behaviours shown towards the person chosen first, irrespective of the choice being correct or incorrect. Rhombus indicates the mean. The thick horizontal black line passing through the boxplots indicates median and the dots indicate the outliers.

### 3.7. Latency-1 for correct versus incorrect choice

Latency-1 did not differ significantly between the dogs that chose the correct person compared to the dogs that chose the incorrect person on any of the days from Day 2 to 5 (Mann Whitney U test, Fig 5a). Since on Day 1, the experimental paradigm required the dogs to choose the incorrect person first, latency-1 for Day 1 was not included in the analysis. For details on the *P* values, check Supplementary Table 5.

### 3.8. SI-1 for correct versus incorrect choice

Across Days 2 to 5, the SI-1 was significantly higher for the dogs that chose the correct person compared to the dogs that chose the incorrect person only on Day 5 (Mann Whitney U test: W=102, P=0.020, Fig 5b). Since on Day 1, the experimental paradigm required the dogs to choose the incorrect person first, SI-1 for Day 1 was not included in the analysis. For details on the *P* values, check Supplementary Table 6.

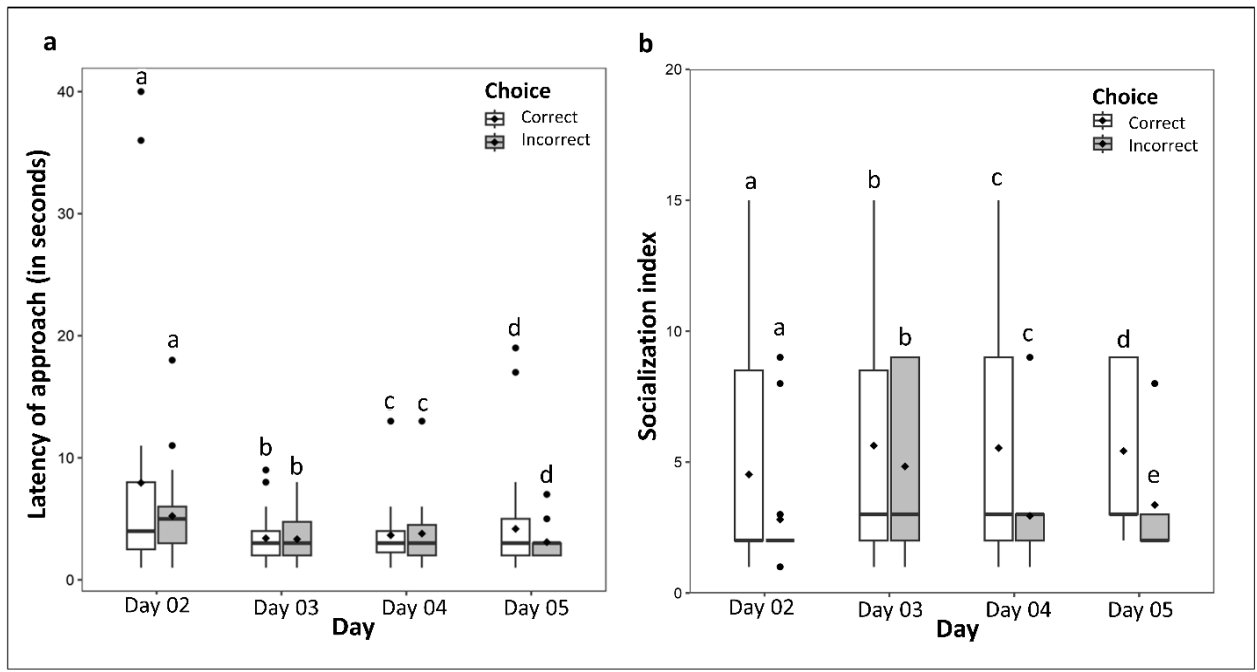

**Fig 5.** A box and whiskers plot showing (a) Latency of approach (in seconds) during first choice (latency-1) for dogs making correct and incorrect choice from Day 2 to 5 (b) SI during first choice (SI-1) for dogs making correct and incorrect choice from Day 2 to 5. The correct person is the one who gave reward as opposed to the incorrect person who did not give reward. Rhombus indicates the mean. The thick horizontal black line passing through the boxplots indicates median and the dots indicate the outliers.

### 3.9. SI during first and second choice

The dogs approached the second person (the one providing reward) only when they approached the incorrect person (the one not providing reward) first. SI for first choice/incorrect choice was calculated based on the behaviours the dog showed towards the person from whom the dog did not receive the reward. SI for second choice/correct choice was defined as the behaviours shown by the dogs towards the person from whom they received the reward. SI during first and second choice was compared only for dogs that made an incorrect choice on a given day. On Day 1, when the dogs were completely unaware of the person who rewarded them, the SI did not differ significantly between the correct and the incorrect choice (Wilcoxon signed rank test: V=18, $P$=0.137, Supplementary Fig 3). SI was significantly higher for the second choice/correct choice only on Day 4 (Wilcoxon signed rank test: V=45, $P$=0.005) and Day 5 (Wilcoxon signed rank test: V=21, $P$=0.03). For details on the $P$ values, check Supplementary Table 7.

### 3.10. Latency-1 on the test day

The latency of approach did not differ between dogs that chose the correct person versus the incorrect person on the test day (Mann-Whitney U test: W = 249, $P$ = 0.298, Fig 6a).

### 3.11. SI-1 on the test day

On the test day (Day 6), when none of the experimenters had food with them, the SI-1 shown by free-ranging dogs was significantly higher when the correct person was chosen as compared to when the incorrect person was chosen (Mann-Whitney U test: W = 101.5, $P$ = 0.005, Fig 6b).

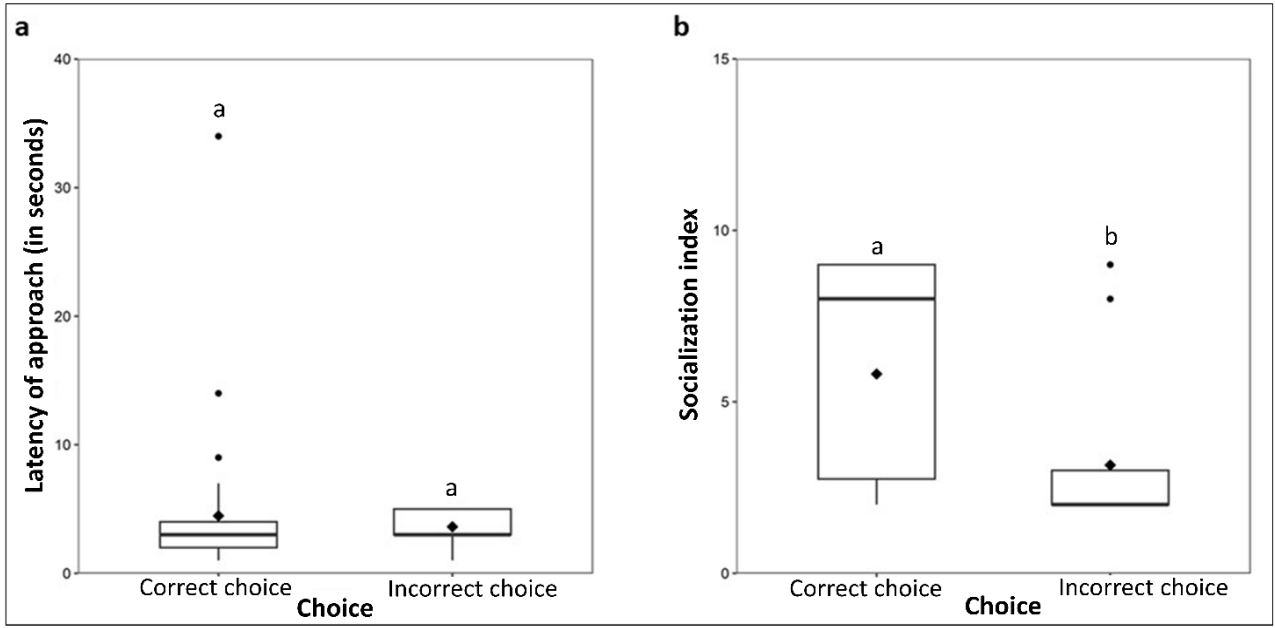

**Fig 6.** A box and whiskers plot showing (a) Latency of approach (in seconds) during first choice (latency-1) for dogs making correct and incorrect choice on Day 6 (test day) (b) SI during first choice (SI-1) for dogs making correct and incorrect choice on Day 6 (test day). The correct person is the one who gave reward as opposed to the incorrect person who did not give reward. Rhombus indicates the mean. The thick horizontal black line passing through the boxplots indicates median and the dots indicate the outliers.

4. Discussion

Faster response and choosing the target/correct choice over time are typical signatures of learning in experiments involving repeated trials with animals. In our experiment, the free-ranging dogs made a higher percentage of correct choice on the 5$^{th}$ and 6$^{th}$ days of the experiment, as compared to Day 2. This suggests that a minimum of four interactions, each consisting of ~20s, with unfamiliar humans is sufficient for learning about the identity of the reward giver in free-ranging dogs. Male and female free-ranging dogs did not differ in their tendency to approach and interact with an unfamiliar person during the approachability test. Only dogs that were not mating, pregnant or lactating were selected for this experiment. We will need further experiments to understand if there would be difference among the two sexes if trials are conducted with individuals in these life history stages. There was no difference in the percentage of dogs that chose the two experimenters on Day 1, suggesting that there was no inherent bias towards any of the two experimenters for the dogs tested.

Free-ranging dogs exhibited an increase in their tendency to approach (depicted by the decline in latency of approach of first choice or latency-1) from Day 1 to Day 2. The tendency to approach (latency-1) however reached saturation thereafter. This increase in tendency to approach was also associated with an increase in the socialization index for the first person chosen or SI-1. The socialization index (SI-1) however showed a significant increase on Day 3 compared to Day 1. This is in line with the findings of Bhattacharjee et al. (2017) who stated that although a single positive social interaction with humans is sufficient to elicit a higher tendency to approach, it does not translate into a higher tendency to interact with unfamiliar humans. In the present study, both food and petting were used as rewards for the dogs. This is in contrary to the previous studies conducted on this line by Bhattacharjee et al. (2017) where the authors used either of the two rewards and found that petting is more efficient as a reward for developing trust towards unfamiliar humans.

For the dogs that made an incorrect choice on the first chance, the tendency to interact was higher for the correct person compared to the incorrect person only on Days 4 and 5. This happened probably because the dogs took a few days to learn about the experimental paradigm which required them to approach the second person when they were denied reward from the person chosen first. Between the dogs that chose the correct versus incorrect person on the first chance, the tendency to interact was higher for the correct person compared to the incorrect person only on Day 5, indicating that the dogs took a few interactions across a few days to distinguish between the two persons, one giving them reward and the other not. One of the possible explanations for this might be that once the dogs reached close to the experimenter, they would have been able to assess whether the person chosen was the reward giver, based on their olfactory memories.

On the test day, free-ranging dogs making correct and incorrect choices did not differ in their latency of approach. However, dogs that chose the correct person showed a higher tendency to interact with the person compared to dogs that chose the incorrect person. These two findings in concert suggests that free-ranging dogs are not very good at distinguishing between humans from a distance of greater than three meters (Polgár et al., 2015). However, when they come close to the experimenters, they are able to recognize them with greater accuracy, leading to the observed difference in their tendency to interact with the correct and the incorrect persons.

Dogs are territorial animals, and free-ranging dogs are known to maintain and defend their territories actively. They routinely encounter familiar as well as unfamiliar humans, and are exposed to a varied range of interactions from humans. Humans are their primary source of food and the free-ranging dogs are not only efficient scavengers, but also adept at begging from humans (Sen Majumder et al., 2014; Finzi et al., 2023). Some people actively feed dogs on a regular basis, while many respond to begging gestures from dogs if they have food with them. Hence, it would be adaptive for the dogs to remember humans who provide them with food, and/or show them love through petting. It would be interesting to check in the future if free-ranging dogs prefer to beg from people who provide them with rewards in earlier interactions, as opposed to those who do not. Female dogs prefer to den near human habitation, utilizing human provided resources (Paul et al., 2016a; Sen Majumder et al., 2016). The ability of individual human recognition will also be of significance in this context as denning near houses of people providing them reward will be beneficial both in terms of higher access to resources and safety. People who provide dogs with social reward are less likely to harm them and their offspring.

In this study, we found that the free-ranging dogs need around four interactions over four days to learn the identity of the human who provides them with rewards, and they use this recognition on consecutive days to their benefit. Thus, it is likely that this ability to recognize and remember individual humans is a faculty of dogs that does not require extensive training or association with individual humans, as in the case of pet dogs. Gray wolves, the closest species to the domestic dog, have been shown to distinguish between familiar and unfamiliar humans by their voice (Gammino et al., 2023). It is likely that dogs evolved with this innate ability to distinguish between humans, which would have aided them in developing close bonds with humans during the process of domestication. They perhaps use a combination of auditory, visual and olfactory cues for this process. Further studies need to be conducted to identify the exact process of human recognition, and the cues used by the dogs for this process, and to understand how long the free-ranging dogs retain the memory of a friendly human without further reinforcement.

**Ethical considerations**

Only non-invasive procedures were used in the experiments. The protocol followed was very similar to the method used by people on the streets of West Bengal to call and interact with the free-ranging dogs and therefore elicited minimum disturbance to the animals studied.

**Funding information**

This research was supported by the Janaki Ammal – National Women Bioscientist Award, Department of Biotechnology. S.N. was supported by the INSPIRE Fellowship, Department of Science and Technology. M.C. was supported by the IISER Kolkata Institute Fellowship.

**CRediT authorship contribution statement**

**Srijaya Nandi:** Conceptualization, Methodology, Formal analysis, Investigation, Data curation, Writing – Original Draft, Visualization, Project administration. **Mousumi Chakraborty:** Investigation. **Aesha Lahiri:** Investigation. **Hindolii Gope:** Investigation. **Sujata Khan Bhaduri:**

Investigation. **Anindita Bhadra:** Conceptualization, Methodology, Resources, Writing – Review & Editing, Supervision, Funding acquisition.

# Declaration of Competing Interest

The authors declare that they have no known competing financial interests or personal relationships that could have appeared to influence the work reported in this paper.

# Acknowledgements

We express our sincere gratitude to Ms. Paramita Paul, Dr. Rubina Mondal, Dr. Udipta Chakraborty and Mr. Rohan Sarkar for their valuable suggestion on data visualization and analysis. We also thank Mr. Arpan Bhattacharyya for designing the illustration given in Fig 1.

**Supplementary material**

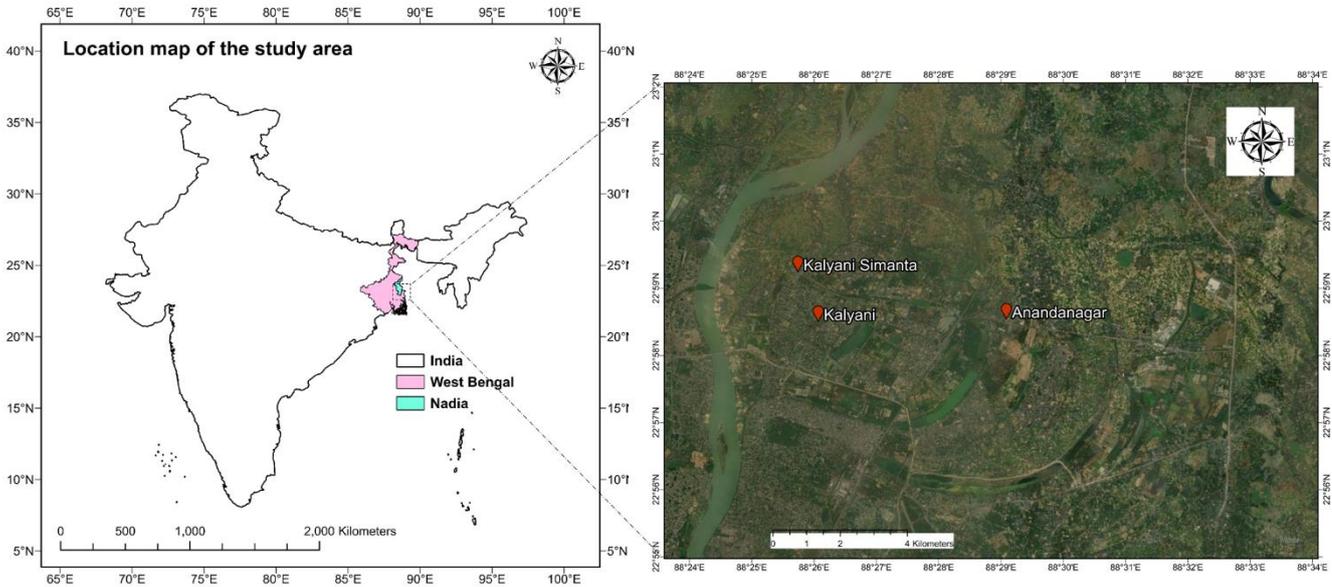

**Fig 1.** Study sites

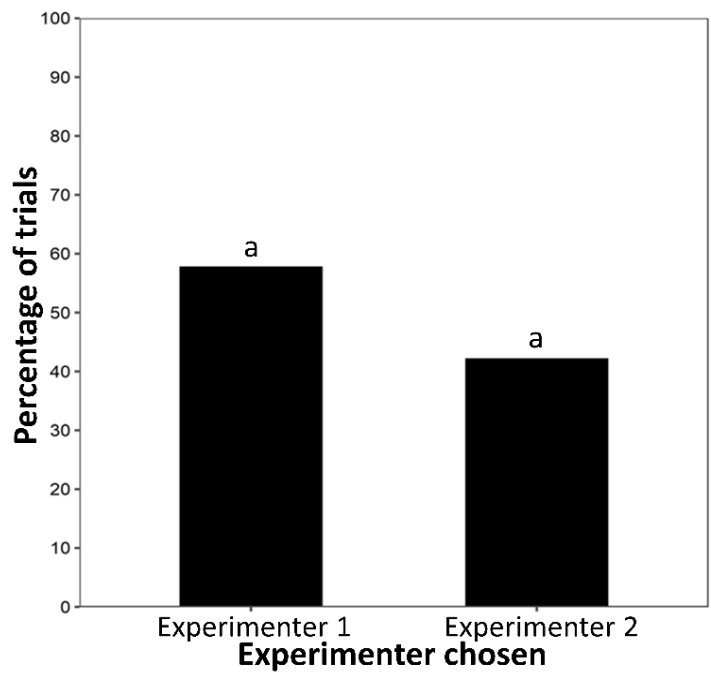

**Fig 2.** A bar graph showing the percentage of trials in which the two experimenters were chosen. There was no significant difference in the preference of dogs for the two experimenters (Chi-square goodness-of-fit test: $\chi^2$=1.0889, df=1, *P*=0.297).

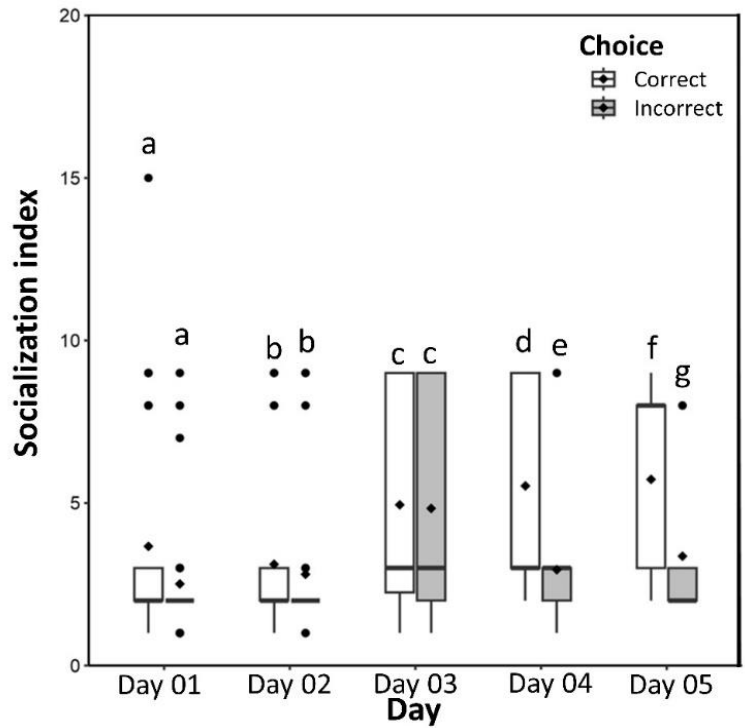

**Fig 3.** Socialization index (SI) for correct person (second choice) versus incorrect person (first choice) for dogs that made an incorrect choice first. Boxplot showing the socialization index (SI) for the correct person (correct choice) and the incorrect person (incorrect choice). Rhombus indicates the mean SI. The thick horizontal black line passing through the boxplots indicates median latency and the dots indicate the outliers. SI was significantly higher for the second/correct person compared to the first/incorrect person only on Day 4 and Day 5.

| Day | χ2 | P |
|---|---|---|
| 2 | 1.089 | 0.297 |
| 3 | 1.8 | 1.179 |
| 4 | 1.089 | 0.297 |
| 5 | 11.756 | 0.0006 |
| 6 | 8.022 | 0.005 |

**Table 1.** Results of the Goodness of fit Chi-square test conducted between correct and incorrect choice from Day 2 to Day 6.

|  | Day 03 | Day 04 | Day 05 | Day 06 |
|---|---|---|---|---|
| **Day 02** | 0.139 | 0.206 | 0.003 | 0.011 |

**Table 2.** *P* for the post-hoc test with Bonferroni correction for comparison between the correct choice between Day 2 and all the other days (Day 3 to 6).

|  | Day 1 | Day 2 | Day 3 | Day 4 | Day 5 |
|---|---|---|---|---|---|
| **Day 2** | 0.022 | - | - | - | - |
| **Day 3** | <0.0001 | 0.193 | - | - | - |
| **Day 4** | <0.0001 | 0.625 | 1.000 | - | - |
| **Day 5** | <0.0001 | 0.374 | 1.000 | 1.000 | - |
| **Day 6** | <0.0001 | 0.514 | 1.000 | 1.000 | 1.000 |

**Table 3.** *P* for the Wilcoxon rank sum test with Bonferroni correction conducted for pairwise comparison of latency of approach of first choice (latency-1) across days.

|       | Day 1   | Day 2  | Day 3 | Day 4 | Day 5 |
|-------|---------|--------|-------|-------|-------|
| Day 2 | 1.000   | -      | -     | -     | -     |
| Day 3 | 0.0004  | 0.1177 | -     | -     | -     |
| Day 4 | 0.0006  | 0.1341 | 1.000 | -     | -     |
| Day 5 | <0.0001 | 0.015  | 1.000 | 1.000 | -     |
| Day 6 | <0.0001 | 0.051  | 1.000 | 1.000 | 1.000 |

**Table 4.** *P* for the Wilcoxon rank sum test with Bonferroni correction conducted for pairwise comparison of socialization index (SI) of first choice (SI-1) across days.

| Day | W     | *P*   |
|-----|-------|-------|
| 2   | 256.5 | 0.835 |
| 3   | 239.5 | 0.943 |
| 4   | 246.0 | 0.991 |
| 5   | 169.0 | 0.653 |

**Table 5.** Results of the Mann-Whitney U test conducted for comparison of latency-1 for correct versus incorrect choice across days.

| Day | W | P |
| --- | --- | --- |
| 2 | 190.5 | 0.139 |
| 3 | 235.5 | 0.868 |
| 4 | 195.5 | 0.216 |
| 5 | 102 | 0.020 |

**Table 6.** Results of the Mann-Whitney U test conducted for comparison of SI-1 for correct versus incorrect choice across days.

| Day | V | P |
| --- | --- | --- |
| 1 | 18 | 0.137 |
| 2 | 6 | 0.174 |
| 3 | 12 | 0.829 |
| 4 | 45 | 0.005 |
| 5 | 21 | 0.03 |

**Table 7.** Results of the Wilcoxon signed rank test conducted for pairwise comparison of socialization index (SI) between the correct and the incorrect choice across days.

**Video 1**

https://drive.google.com/file/d/1LvRFrMG7pfEmG_THDKHnQg2bwKz-Wh6e/view?usp=drive_link

**Video 2**

https://drive.google.com/file/d/1RyWnRhxANpBAZYR6GOIFDK8IOGVeIsM-/view?usp=drive_link

**Video 3**

https://drive.google.com/file/d/1kb5ars8hSsfRKo9IXAD-ZdxpDEWjjeFA/view?usp=drive_link